# OPTIMIZE THE PARAMETERS OF THE PID CONTROLLER USING GENETIC ALGORITHM FOR ROBOT MANIPULATORS
TỐI ƯU THAM SỐ BỘ ĐIỀU KHIỂN PID SỬ DỤNG THUẬT TOÁN DI TRUYỀN CHO ROBOT

*Vu Ngoc Son[1], Pham Van Cuong[1], Nguyen Duy Minh[2], Phi Hoang Nha[1*]*
[1] *Falcuty of Electrical Engineering, Hanoi university of Industry*
[2] *Falcuty of Mechanical Engineering, Hanoi university of Industry*

**ABSTRACT**

This paper presents the design a Proportional-Integral-Derivative (PID) controller with optimized parameters for a two-degree-of-freedom robotic arm. A genetic algorithm (GA) is proposed to optimize the controller parameters, addressing the challenges in determining PID controller parameters for highly nonlinear systems like robotic arms compared to traditional methods. The GA-optimized PID controller significantly improves control accuracy and performance over traditional control methods. Simulation results demonstrate that the robotic arm system operates with high precision and stability. Additionally, the shortened trajectory tracking response time enhances the feasibility of applying this control algorithm in real-world scenarios. This research not only confirms the suitability of PID-GA for robotic arms and similar systems but also opens new avenues for applying this algorithm to real physical systems.

**Tóm Tắt:**

Bài báo trình bày thiết kế bộ điều khiển PID với tham số được tối ưu cho tay máy robot hai bậc tự do. Thuật toán di truyền (GA) được đề xuất nhằm tối ưu tham số cho bộ điều khiển, giúp khắc phục những hạn chế trong việc khó xác định tham số bộ điều khiển PID cho đối tượng có tính phi tuyến mạnh như tay máy robot so với các phương pháp xác định tham số điều khiển truyền thống. Thuật toán tối ưu GA đã cải tiến bộ điều khiển PID có thể điều khiển chính xác vị trí của robot với chất lượng điều khiển được cải thiện đáng kể so với các phương pháp điều khiển truyền thống. Kết quả mô phỏng cho thấy hệ thống tay máy robot hoạt động chính xác, ổn định. Đồng thời, đáp ứng bám quỹ đạo được rút ngắn là một đặc điểm giúp thuật toán điều khiển này trở lên khả thi trong bài toán áp dụng thực tế. Nghiên cứu này không chỉ khẳng định tính phù hợp của PID-GA đối với robot và các đối tượng tương tự, mà còn mở ra hướng nghiên cứu mới trong việc áp dụng thuật toán này trên các hệ vật lý thực tế.

***Từ khóa:*** *Thuật toán di truyền (GA); bộ điều khiển PID; Tay máy robot; Tay máy robot 2 bậc; Điều khiển tối ưu*

## 1. INTRODUCTION

Robotic manipulators play a crucial role in modern manufacturing processes, providing material handling and manipulation capabilities. These devices are designed to operate in diverse environments and employ various motion control techniques to accomplish their tasks. Managing the motion of robotic manipulators becomes increasingly challenging as the degrees of freedom (DOF) increase. Their multivariable nonlinear nature, coupled with uncertainties in dynamics such as non-linear friction, external disturbances, time-varying dynamics, and loads, can adversely affect performance and stability. Thus, ensuring stable and high-performance operation requires understanding, controlling, and monitoring the trajectory of robotic manipulators.

To address these problems, several motion control techniques, including PID control, optimal control, adaptive control, robust control, and neural network control, have been developed [1-8]. Each technique has its strengths and limitations, tailored to specific applications and system requirements. Monitoring the performance of robotic manipulators is equally vital, and sensors like encoders, force/torque sensors, and accelerometers provide valuable feedback for adapting the controller strategy to changing conditions. Additionally, machine learning approaches, such as reinforcement learning, can optimize the control strategy for robotic manipulators.

Among these techniques, the Proportional-Integral-Derivative (PID) control method [9,10] stands out for its simplicity and flexibility, making it widely used in industrial processes and robotic manipulators. PID algorithms are often implemented independently for each joint, regulating various aspects such as movement direction and robot position. Its reliability, cost-effectiveness, ease of implementation, and hardware independence make PID one of the most versatile control methods for mechanical systems. However, designing a PID controller for nonlinear objects like a robotic arm with uncertain parameters poses challenges. Accurate and optimal determination of Kp, Ki, and Kd parameters without sufficient knowledge of the robotic arm system can be difficult. In such cases, adaptive control methods considering changing system dynamics over time or hybrid PID-based control methods utilizing neural networks, genetic algorithms (GA_PID), or particle swarm optimization algorithms (PSO_PID) can optimize control while maintaining stability and performance [11-15].

In this research, we propose a Genetic Algorithm (GA) to determine optimal PID controller parameters for robot manipulators. By employing the GA approach, we aim to overcome the limitations of traditional trial-and-error methods that rely on extensive experience and time-consuming iterations. The GA_PID is developed to select suitable Kp, Ki, and Kd values, ensuring stable and efficient operation of the robot manipulators. Based on simulation results compared with existing results, our proposed controller is more flexible and tracking errors are improved.

This paper is organized as follows. The methods are described in Section 2. Section 3 presents the system description. Section 4 provides simulation results for two-link robot manipulators. Finally, the conclusion is given in Section 5.

## 2. METHODS

The dynamics of two-link robot manipulators were modeled using a set of nonlinear, second-order, ordinary differential equations with the explicit purpose of obtaining accurate and effective control of the robot arm in order to perform various tasks by following specified trajectories within an expected workspace. To simulate the dynamics accurately, the Lagrange-Euler methods were adopted for formulating the mathematical model. For successful implementation of the control strategy investigated based on the derived dynamics equations, the mathematical model must reflect the true representation of the dynamic behavior of the robot arm. In this regard, the modeling of the mathematical equations for the robotic manipulator has been considered critical in this research. The proposed mathematical model reflects the true mechanical properties of the system, including motor torque inputs and links' lengths.

To achieve effective control of the robot manipulators, a PID controller was developed by expanding on the derived mathematical model that comprised several parameters such as motor angle, arm velocity, joint positions, etc. The PID controller was designed to minimize the tracking error and enhance the position control system performance by tuning the PID controller based on the trial-and-error method and refining the initial PID coefficients iteratively until they sufficiently and effectively controlled the robot motion. Moreover, to automate and optimize the PID coefficients, Genetic Algorithms (GAs) have been incorporated into the design process to estimate certain PID controller parameters. GA is utilized to determine the adjustable gains of the PID controller through alterations of the coefficients of each adjustment parameter based on a fitness function.

**2.1 PID Controller**

Basically, PID is combination of three basic building blocks (Fig. 1).

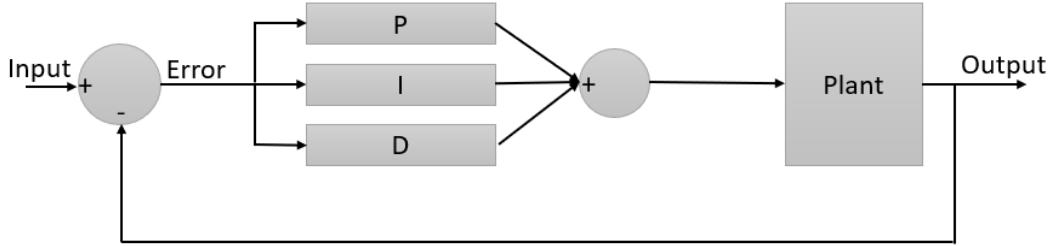

*Fig. 1. Schematic of the PID controller*

In proportional control:
$$P = K_P * \text{Error} \tag{1}$$

It uses proportion of the system error to control the system. In this action, an offset is introduced in the system.

In Integral control:
$$I = K_I * \int \text{Error}.dt \tag{2}$$

It is proportional to the amount of error in the system. In this action, the I-action will introduce a lag in the system. This will eliminate the offset that was introduced earlier on by the Paction.

In Derivative control:
$$D = K_D * \frac{d(\text{Error})}{dt} \tag{3}$$

The influence of the D-action is directly related to the rate of change of the error. By implementing the D-action, the system experiences a leading effect, effectively eliminating any preceding lag introduced by the I-action. These three controllers, when combined, can be described by the following transfer function.

The PID controller operates on the controlled variable through a balanced integration of the P, I, and D control actions. The P component is proportionate to the error signal, defined as the difference between the input and feedback signals. The I component is proportional to the integral of the error signal, while the D component is proportional to its derivative. Integrating these actions enables the realization of a continuous PID controller, widely employed across industries worldwide. Substantial research, studies, and applications have emerged in recent years.

However, a significant challenge lies in determining the precise values of proportional, integral, and derivative gains, particularly when classical methods necessitate comprehensive mathematical modeling of the entire process plant. Classical control approaches, such as the Ziegler-Nichols method, may not yield satisfactory results in scenarios with numerous input/output disturbances or nonlinear plant behavior.

**2.2 Genetic Algorithms**

A chromosomal population was initialized with control parameters tailored for the PID controller. Employing a fitness evaluation function aligned with our predetermined objectives and specifications, a methodical assessment of the chromosomes ensued. Through iterative application, the operations of selection, crossing, and mutation were employed to incrementally refine their fitness and systematically explore innovative combinations of parameters specific to the PID controller. Following each iteration, a thorough comparative analysis was conducted between the offspring and their predecessor generation, with only the most exemplary individuals retained.

This cyclic process persevered until a predetermined maximum number of generations was reached or a fitness level deemed satisfactory was attained. The conclusion of this methodology yielded an optimal collection of control parameters for the PID controller, encapsulated within the solution perceived as the most superior in the context of our study.

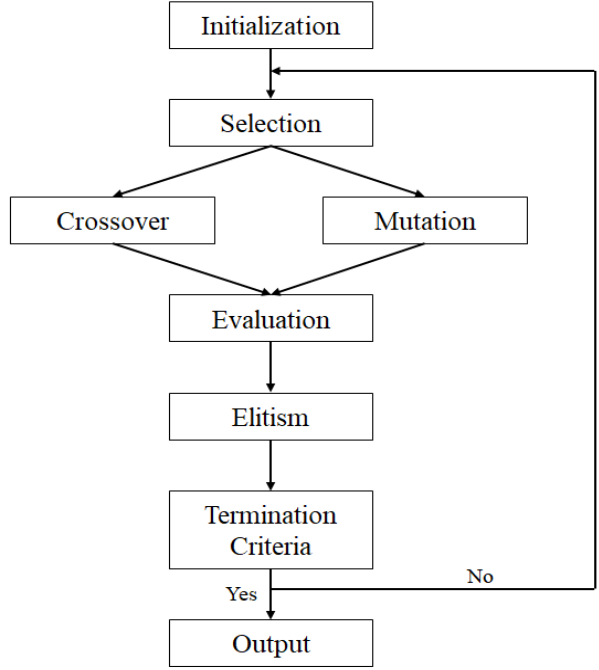

*Fig. 2. Flowchart illustrating the Genetic Algorithm Process*

## 3. SYSTEM DESCRIPTION
### 3.1 Dynamic of robot manipulators

We consider the dynamics of an n-link robot manipulator with external disturbance can be expressed in the Lagrange as follows [1]:

$$M(q)\ddot{q} + C(q,\dot{q})\dot{q} + G(q) + F(\dot{q}) = \tau \qquad (4)$$

where: $(q, \dot{q}, \ddot{q}) \in R^{n \times 1}$ are the vectors of joint position, velocity, and acceleration, respectively. $M(q) \in R^{n \times n}$ is the symmetric inertial matrix. $C(q, \dot{q}) \in R^{n \times n}$ is the vector of Coriolis and centripetal forces. $G(q) \in R^{1 \times 1}$ expresses the gravity vector. $F(\dot{q})$ represents the vector of the frictions. $\tau$ is the joints torque input vector.

### 3.2 Optimizing PID controller parameters using genetic algorithm

To obtain the optimal values for the Kp, Ki, and Kd parameters, we utilize a genetic algorithm (GA). The GA involves encoding the Kp, Ki, and Kd parameters as genes within each chromosome or individual. Specifically, these parameters are encoded into a string of real numbers that represents the individual. An adaptive function called the standardized squared integral error (ISE) is chosen to determine the fitness of each individual in the population.

$$fitess = \int_0^\infty (e_1^2(t) + e_2^2(t))dt \qquad (5)$$

The structure of Genetic algorithm-PID is shown in figure 3

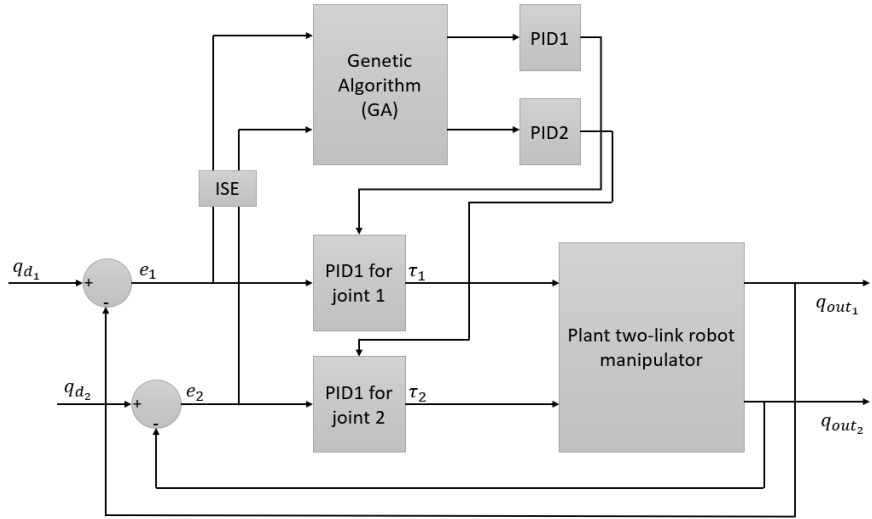

*Fig. 3. Flowchart illustrating the Genetic Algorithm Process*

## 4. SIMULATION RESULTS
### 4.1 Dynamic of a two-link robot manipulator

n this section, a two-link robot manipulator is utilized to verify the effectiveness of the proposed control scheme and a series of simulation research is carried out.

We consider the two-link robot manipulator model that is shown in Fig. 4, and the dynamic equation can be described by using Lagrange method as (4):

$$M(q)\ddot{q} + C(q,\dot{q})\dot{q} + G(q) + F(\dot{q}) = \tau$$

where:

$$M(q) = \begin{bmatrix} (m_1+m_2)l_1^2 + m_2l_2^2 + 2m_2l_1l_2\cos(q_2) & m_2l_2^2 + m_2l_1l_2\cos(q_2) \\ m_2l_2^2 + m_2l_1l_2\cos(q_2) & m_2l_2^2 \end{bmatrix}$$

$$C(q,\dot{q}) = \begin{bmatrix} -m_2l_1l_2\sin(q_2)(2\dot{q}_1\dot{q}_2 + \dot{q}_2^2) \\ -m_2l_1l_2\sin(q_2)\dot{q}_1\dot{q}_2 \end{bmatrix}$$

$$C(q,\dot{q}) = \begin{bmatrix} -(m_1+m_2)l_1 g\sin(q_1) - m_2l_2 g\sin(q_1+q_2) \\ -m_2l_2 g\sin(q_1+q_2) \end{bmatrix} \quad \tau = \begin{bmatrix} \tau_1 \\ \tau_2 \end{bmatrix};$$

in which $m_1$ and $m_2$ are the mass of link 1 and link 2, respectively, $l_1$ and $l_2$ are the length of link 1 and link 2, respectively, and g is acceleration of gravity. The parameters of two-link robot manipulator are given as: $m_1 = 5kg$; $m_2 = 5kg$; $l_1 = 0,34m$; $l_2 = 0,34m$; $g = 9,81m/s^2$ [9]

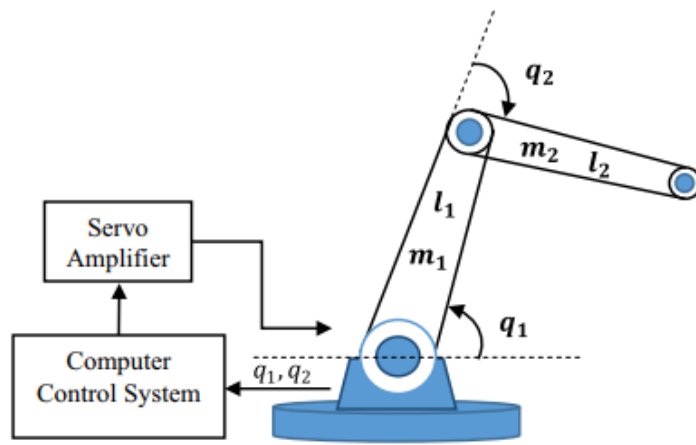

*Fig. 4. Two-link robot manipulator control system [1]*

The control objective is to control the joint angles of a two-link robot manipulator to follow the desired trajectories. The desired position trajectories of two-link robot manipulators are chosen by: $\begin{bmatrix} q_{d1} \\ q_{d2} \end{bmatrix} = \begin{bmatrix} \frac{\pi}{2} \\ \pi \end{bmatrix}$ and initial: $\begin{bmatrix} q_{01} \\ q_{02} \end{bmatrix} = \begin{bmatrix} \pi \\ \frac{\pi}{2} \end{bmatrix}$

### 4.2 Parameters Ki, Kp and Kd by genetic algorithm (GA)

The simulation is performed with a maximum number of generations of 1000. The number of parents used in the population is 20. The coefficient of hybridization in the population is 0.6. The coefficient of mutation in the population is 0.4.

The algorithm has converged after 82 generations and give the results in table 1:

***Table 1.*** *Table optimized PID parameter using GA algorithm*

| Parameter | $K_{P1}$ | $K_{I1}$ | $K_{D1}$ | $K_{P2}$ | $K_{I2}$ | $K_{D2}$ | Fitness |
|---|---|---|---|---|---|---|---|
| Value | 97.4700 | 97.4700 | 97.4700 | 97.4700 | 97.4700 | 97.4700 | 97.4700 |

### 4.3 The simulation results

In this section, the simulation results obtained from simulating the PID (Proportional-Integral-Derivative) controller for a twolink robot manipulators are presented. The PID parameters were used based on the controller's experience [9]. Additionally, a set of optimized parameters generated through a genetic algorithm (Table 1) were employed, and their performance was compared.

- Parameters of PID1 [9]: $K_{P1}$=30, $K_{I1}$=20, $K_{D1}$=12, $K_{P2}$=32, $K_{I2}$=30, $K_{D2}$=22.
- Parameters of PID by GA: $K_{P1}$=97.4700, $K_{I1}$=98.0500, $K_{D1}$=13.4600, $K_{P2}$=98.5200, $K_{I2}$=70.2400, $K_{D2}$=12.1500.

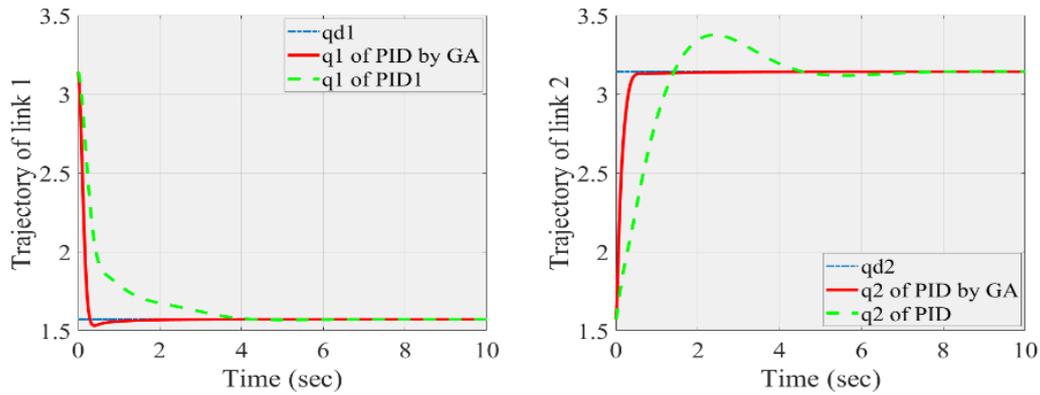

***Fig. 5.*** *Simulation results of trajectory of 2-link*

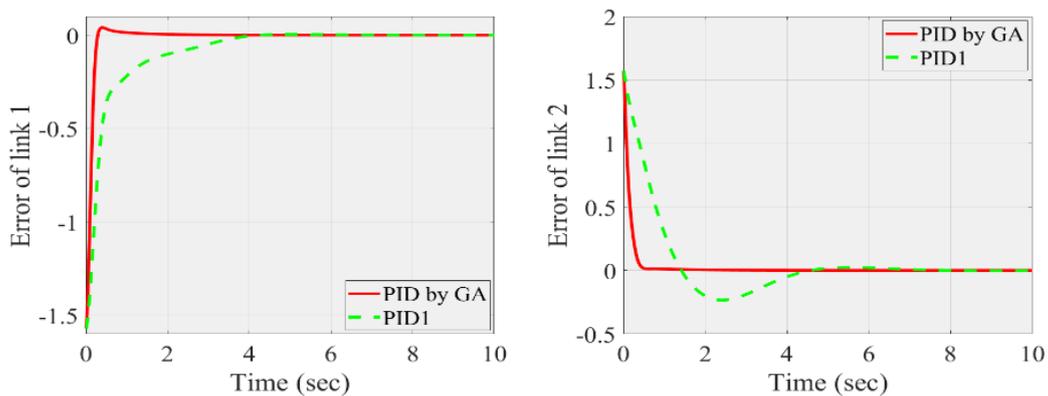

***Fig. 6.*** *Simulation results of error of 2-link*

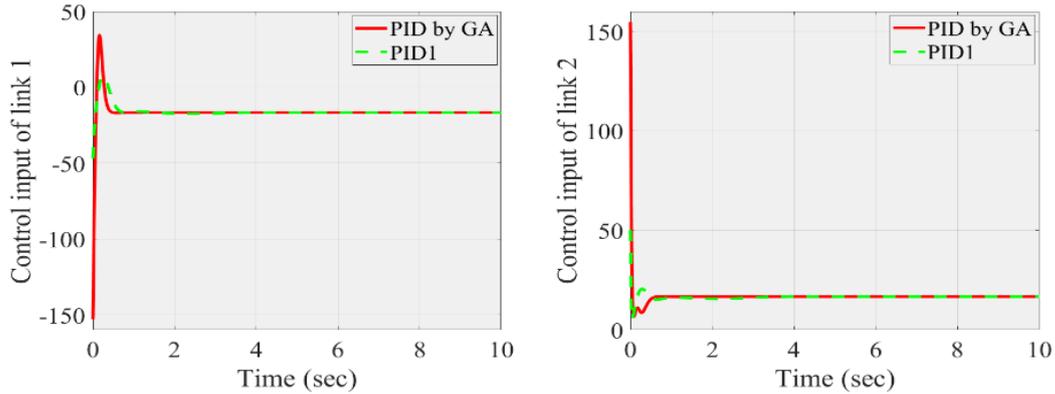

***Fig. 7.*** *Simulation results of control input of 2-link*

The superiority of the genetically optimized PID controller compared with the empirically proposed PID parameter set is demonstrated by the simulation results. It is observed that the genetically optimized PID controller surpasses the empirically proposed one in various quality criteria. The responses of the system controlled by the genetically optimized PID showcase reduced overshoot, faster settling time, and improved steady-state error, effectively meeting the desired control objectives. This emphasizes the significance of evaluating controller performance based on multiple quality standards.

Additionally, enhanced stability and robustness are exhibited by the genetically optimized PID controller during the simulation. It successfully suppresses oscillations, resulting in smoother control actions and minimizing the risk of system instability. This indicates that the proposed controller is designed with a comprehensive understanding of the system dynamics and possesses the capability to handle disturbances and uncertainties more effectively. Consequently, it aligns with the highest quality standards in control engineering.

## 5. CONCLUSIONS

In this paper, we have successfully investigated using a PID controller to regulate robot manipulators. The simulation results demonstrated that by employing a Genetic Algorithm to optimize parameters for the PID controller, the performance of the control system in terms of adherence and deviation significantly improved over conventional PID control techniques. The study showed that the proposed GA-based approach is an efficient method for determining PID control parameters that can be applied to robotic systems. Our approach provides an optimal solution that ensures precise tracking performance with minimum overshoot and settling time. Furthermore, we expect our research results will increase the quality of controllers with optimized parameters working in complex robotic control systems. As such, it opens up the potential for exploring new frontiers in advanced robotics control. Moreover, we anticipate extending this approach to control more complex robot manipulators and other types of systems.